\documentclass[11pt]{article}
\usepackage[varbb]{newpxmath}
\usepackage{graphicx} 
\usepackage[margin=0.8in]{geometry}
\usepackage{hyperref}
\usepackage[
backend=biber,
style=ieee,
]{biblatex}

\newlength{\bibparskip}
\let\oldthebibliography\thebibliography
\renewcommand\thebibliography[1]{%
  \oldthebibliography{#1}%
  \setlength{\parskip}{\bibitemsep}%
  \setlength{\itemsep}{\bibparskip}%
}

\AtBeginBibliography{\footnotesize}

\title{Quantum abstract machines without circuits: the need for higher algorithmic expressiveness}
\author{Santiago Núñez-Corrales${}^{1,2}$ -- \texttt{nunezco2@illinois.edu}  \\
${}^{1}$National Center for Supercomputing Applications\\
${}^{2}$Illinois Quantum Information Science and Technology Center\\
University of Illinois Urbana-Champaign\\
}
\date{}

\begin{document}

\maketitle

\begin{abstract}
    Existing abstract models of quantum computation make reference to circuit elements, much in contrast to their classical counterparts. Circuits, as a model of computation, substantially limit algorithmic expression and obscure high-level connections between problems and quantum resources. It is argued here that new models are needed to achieve high-level algorithmic expressiveness that allow composable procedural abstractions to manifest, leading to the development of instructions in the sense usually understood in high-level programming languages. Doing so appears essential to the discovery of new quantum algorithms, and deeper understanding of how quantum resources compose into useful patterns, or \emph{quantum motifs}. To achieve this, stronger investment in the intersection between higher-algebra, mathematical physics and quantum science is required to cope with future challenges brought forth by \textit{very large quantum scale integration}.
\end{abstract}

\vspace{5pt}

\textbf{Topics:} foundations of quantum computation; design and analysis of established and novel abstract quantum computing and programming models; ASCR Workshop on Quantum Computing and Networking
\vspace{5pt}

\textbf{Challenge:} Defining what computation means requires understanding what resources are available and how these can be marshalled to accomplish tasks we identify as computing at an abstract level. Having acquired this knowledge in advance partially explains the successful evolution of classical computing technologies, in which universal Turing machines and lambda calculus delineated hardware requirements sufficiently well. Later these were satisfied through the match between bi-stable digital electronics and Boolean algebra which resulted in finite models of arithmetic in use today. To reiterate: the core theoretical models of computation via abstract machines were readily available well in advance of the transistor, which made algorithmic development agnostic to hardware details even at VLSI scales. We characterize here the information flow going from theory to hardware as \textit{top-down}, and originated from questions about provability of theorems in (integer) arithmetic. 

Quantum computing inverts the paradigm described above, going \textit{bottom up}. All formulations of quantum Turing machines, quantum random access machines \cite{knill1996conventions,wang2023quantum}, \cite{deutsch1985quantum,feynman1986quantum,guerrini2020quantum} and quantum $\lambda$-calculus \cite{selinger2009quantum,boto2023zeta} so far prescribe the execution of gates on primitive quantum steps --primitive in the sense of mapping directly to quantum hardware- thus conflating the abstract concept of performing computation with the concrete task of building circuits. Three reasons seem to largely explain the latter: (a) the original intent involved modeling physical processes as computation at the most fundamental level (and viceversa), (b) reasoning about small quantum systems is more feasible that for large ones, and (c) building experiments that benefit from decades of research in quantum science was directly attainable.

Despite the ongoing quantum computing revolution and substantial funding dedicated to it, progress the number of quantum algorithms is scant. Going from gates and qubits to a general way to create new ones remains elusive. In a hypothetical situation in which classical computing followed a similar path, defining and understanding computation from the basis of bi-stable elements and Boolean algebra would have been substantially harder; the fact that quantum devices offer a much larger array of resources for computation \cite{chitambar2019quantum} compounds our difficulties even more. To put it bluntly, quantum information --and by extension quantum mechanics \cite{bub2005quantum}- does not provide a sufficient basis to understand quantum computation in abstract terms.

\vspace{5pt}

\textbf{Opportunity:} To reap the expected benefits of quantum computing, abstract machine models capable of facilitating the development of algorithms at higher conceptual levels are direly needed. Regardless of how these are specified --e.g., instructions vs functions- it is clear that outcome of such exercise will produce composable procedural abstractions: entities that operate well beyond Hilbert spaces and their transformations, that can be combined to produce useful generative effects, and that provide a denotational semantics of future quantum programming languages which makes no explicit reference to circuits. In particular, finding an analogue to finite models of integer arithmetic for quantum computing would allow specifying new classes of quantum abstract machines in terms of their (possibly discrete) transformations without worrying about hardware details of any sort, and then delegate hardware details to computer architecture designers and, later, to specialized hardware compilers.

Succinctly: composable procedural abstractions are essential for the sustained development and discovery of new quantum algorithms; these cannot be found amidst quantum circuits. Creating the funding and research context around this challenge will produce a ``quantum jump'' in the way we characterize computational problems \cite{chapman2022quantum}, and consequentially in the number of applications for which quantum computing yields a concrete advantage. 

\vspace{5pt}

\textbf{Assessment:} At present, some of the ingredients needed to develop high-level quantum abstract machines of the sort we look for seem to be present. These include: the relation between Clifford algebras and quantum field theory \cite{baugh2001clifford}, the ability to define abstract machines using geometrical algebra \cite{schott2010reductions}, symmetries present in quantum random walks with implications for quantum automata \cite{arnault2022clifford}, and higher algebra formulations of quantum field theory \cite{halvorson2006algebraic,basti2017quantum,benini2019homotopy}. However, they are insufficient for the task in their present form. It is not immediately obvious how to arrange them to produce new theories and abstract machines directly at a high level, which other ones are missing, or whether the resulting theories will hold once the limits of entanglement and other quantum resources are more deeply explored.

However, preliminary research and practice in quantum programming suggests a possible route: privileging those theories where combinators arise, and where certain patterns in circuit-building can be abstracted away as \emph{quantum motifs} where details such as the number of qubits, specific gates and how these can be optimized disappear. Combinators proved fundamental in early days of classical computing to capture meta-patterns that simplify reasoning about how we construct algorithms and data structures. This work requires investing in interdisciplinary teams that include theoretical physicists, mathematicians, logicians, and computer scientists. While the investment for a single team looks modest in comparison to experimental work in quantum computing, the difficulty of the task calls for sufficient funding across relevant agencies to maximize the overall surface of attack through multiple research teams, thus maximizing the probability of finding good candidate theories and abstract machines.

\vspace{5pt}

\textbf{Timeliness:} Predicted timelines in quantum hardware manufacturers suggest an upcoming era of \emph{very large quantum scale integration} (VLQSI). Devising and implementing large algorithms will become rapidly infeasible for humans. Given the level of investment in quantum computing, the capabilities of devices for theory testing, the increasing pressure to move from proof of principle to applications, and how limited progress on this issue has been achieved, the urgency of this challenge cannot be understated.

\printbibliography
\end{document}